\documentclass[11pt,twoside]{article}
\usepackage{./asp2014}

\aspSuppressVolSlug
\resetcounters

\begin{document}

\title{A study of the stellar photosphere-hydrogen ionisation front interaction in pulsating variables using period-colour relations}
\author{Susmita~Das,$^1$ Shashi M. Kanbur,$^{2}$ Earl P.~Bellinger,$^{3}$ Anupam Bhardwaj,$^{4}$ and Harinder P. Singh,$^{5}$
\affil{$^1$Department of Physics \& Astrophysics, University of Delhi, Delhi 110007, India; \email{susmitadas130@gmail.com}}
\affil{$^2$Department of Physics, State University of New York Oswego, Oswego, NY 13126, USA; \email{shashi.kanbur@oswego.edu}}
\affil{$^3$Stellar Astrophysics Centre, Department of Physics and Astronomy, Aarhus University, Aarhus 8000, Denmark;\\ School of Physics, UNSW Sydney, NSW 2052, Australia; \email{bellinger@phys.au.dk}}
\affil{$^4$Kavli Institute for Astronomy and Astrophysics, Peking University, Yi He Yuan Lu 5, Hai Dian District, Beijing 100871, China; \email{abhardwaj@pku.edu.cn}}
\affil{$^5$Department of Physics \& Astrophysics, University of Delhi, Delhi 110007, India; \email{hpsingh@physics.du.ac.in}}}

\paperauthor{Susmita~Das}{susmitadas130@gmail.com}{https://orcid.org/0000-0003-3679-2428}{University of Delhi}{Department of Physics \& Astrophysics}{Delhi}{Delhi}{110007}{India}
\paperauthor{Shashi M. Kanbur}{shashi.kanbur@oswego.edu}{}{State University of New York Oswego}{Department of Physics}{Oswego}{New York}{13126}{USA}
\paperauthor{Earl P.~Bellinger}{bellinger@phys.au.dk}{}{Aarhus University}{Stellar Astrophysics Centre, Department of Physics and Astronomy}{Aarhus}{Aarhus}{8000}{Denmark}
\paperauthor{Anupam Bhardwaj}{abhardwaj@pku.edu.cn}{}{Peking University}{Kavli Institute for Astronomy and Astrophysics}{Beijing}{Beijing}{100871}{China}
\paperauthor{Harinder P. Singh}{hpsingh@physics.du.ac.in}{}{University of Delhi}{Department of Physics \& Astrophysics}{Delhi}{Delhi}{110007}{India}

\begin{abstract}
Period-colour (PC) relations may be used to study the interaction of the stellar photosphere and the hydrogen ionisation front (HIF). RR~Lyrae (RRL) and long period classical Cepheids ($P>10$d) have been found to exhibit different PC behavior at minimum and maximum light which can be explained by the HIF-photosphere interaction based on their location on the HR diagram. In this work, we extend the study to include type~II Cepheids (T2Cs) with an aim to test the HIF-photosphere interaction theory across a broad spectrum of variable star types. We find W~Vir stars and BL~Her stars to have similar PC relations as those from long period and short period classical Cepheids, respectively. We also use MESA to compute RRL, BL~Her, and classical Cepheid models to study the theoretical HIF-photosphere distance and find the results to be fairly consistent with the HIF-photosphere interaction theory.
\end{abstract}

\section{Introduction}

Classical pulsators like RRLs, T2Cs, and classical Cepheids are important astrophysical objects. They are used as tracers of stellar populations and extragalactic distances, thanks to their well-defined period-luminosity relations \citep{muraveva2015, bhardwaj2016a}. \citet{bhardwaj2014, das2018}, and references therein) studied the PC relations to gain insight into the structure of the outer envelopes of RRLs and classical Cepheids. The PC results remained consistent with different observational datasets: RRLs have flat PC slope at minimum light with a significantly greater PC slope at maximum light while long-period classical Cepheids ($P>10$ d) have a flat PC$_\textrm{max}$ slope with a significant PC$_\textrm{min}$ slope. These results may be explained by the interaction of the stellar photosphere and the HIF \citep{kanbur1996}. The difference in the behaviour of PC relations between RRLs and classical Cepheids may be attributed to their different locations on the Hertzsprung-Russell (HR) diagram- RRLs are hotter and less luminous than classical Cepheids. A lower $L/M$ ratio and/or a hotter effective temperature indicates smaller distance between the HIF and the photosphere or equivalently, a situation with ``engaged'' HIF-photosphere.

\begin{figure*}
\centering
\includegraphics[scale = 0.65]{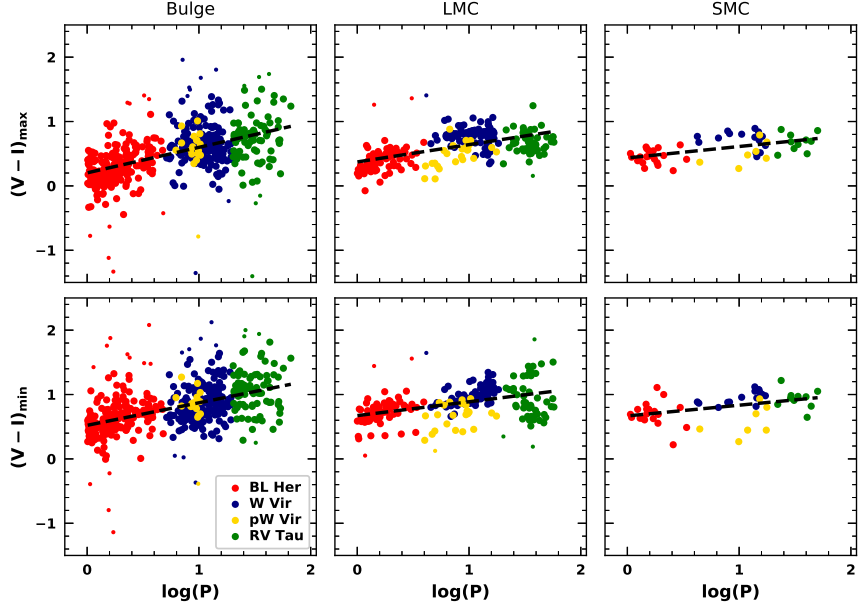}
\caption{The period-colour relations for the different subclasses of type~II Cepheids in the Bulge and the Magellanic clouds at maximum and minimum light. The best fit linear regression for T2Cs as a whole with the different sub-classes combined is represented by the dashed line, after recursively removing the 3$\sigma$ outliers (shown in smaller symbols).}
\label{fig:PC} 
\end{figure*}

Conversely, according to the HIF-photosphere interaction theory, we expect stars occupying similar regions on the HR diagram to show similar behaviour in their PC relations. With an aim to analyse this, we extend our study to include the PC relations of the different subclasses of T2Cs. We use the classification as provided by \citet{soszynski2018}: BL~Hers are those with periods between $1-4$ days, W~Vir stars have periods between $4-20$ days and RV~Tau stars have $P>20$d. We also compute a sample grid of RRL, classical Cepheid, and BL~Her models using the radial stellar pulsation codes in \emph{Modules for Experiments in Stellar Astrophysics} \citep[MESA,][]{paxton2019} to study the distance between the HIF and the photosphere theoretically.

\section{Data and Methodology}

\begin{figure*}
\centering
\includegraphics[scale = 0.5]{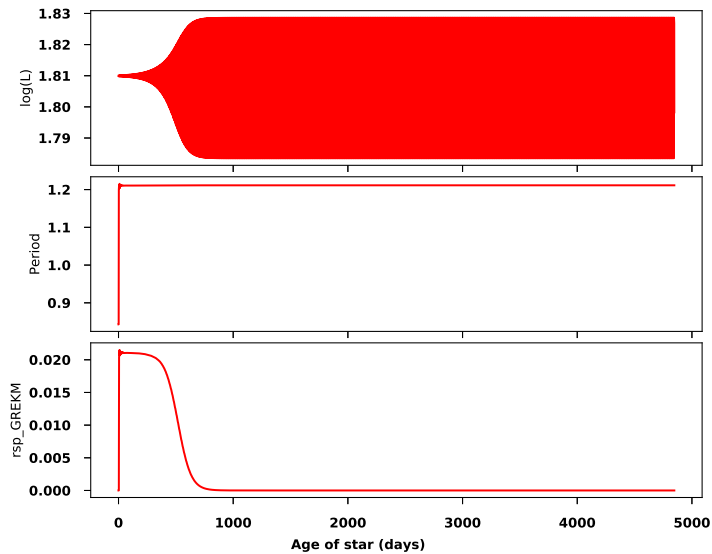}
\caption{An example of a full-amplitude stable pulsation BL~Her model with the input parameters $Z=0.004, X=0.756, M=0.55M_{\odot}, L=64.56L_{\odot}, T=5950$K.}
\label{fig:lc} 
\end{figure*}

The optical ($VI$) light curves of the T2Cs in the Galactic bulge \citep{soszynski2017} and the Magellanic Clouds \citep{soszynski2018} are obtained from the OGLE-IV catalogue. We choose stars with well-sampled light curves for our analysis, having more than 30 epochs of observation in both $V$ and $I$ bands. The light curves are fitted with the Fourier sine series \citep{das2018}: $m = m_0 + \sum_{k=1}^{N}A_k \sin(2 \pi kx+\phi_k)$, where $x$ is the phase and $N$ is the order of fit ($4\leq N\leq 8$) obtained from the Bart's criteria \citep{bart1982}. We define colour at maximum and minimum light as:
\begin{equation}
(V-I)_{\max} = V_{\max} - I_{\operatorname{phmax}};
(V-I)_{\min} = V_{\min} - I_{\operatorname{phmin}},
\end{equation}
where $I_{\operatorname{phmax}}$ and $I_{\operatorname{phmin}}$ correspond to the $I$-mag at the same phase as that of $V_{\max}$ and $V_{\min}$, respectively.

The PC relations for the T2Cs are obtained after correcting for extinction using standard methods. The colour excess for T2Cs in the Magellanic Clouds are obtained from the reddening maps of \citet{haschke2011} while those in the Galactic bulge are obtained from \citet{gonzalez2012}. The extinction values are then estimated by adopting the reddening law of \citet{cardelli1989}. 

\section{Results from the observational aspect}

The PC relations for the T2Cs in the Galactic bulge and the Magellanic Clouds at minimum and maximum light are displayed in Fig.~\ref{fig:PC}. The PC$_\textrm{max}$ slopes for W~Vir stars in the Bulge ($0.037 \pm 0.141$), LMC ($-0.033 \pm 0.086$), and SMC ($-0.169 \pm 0.189$) are not significantly different from zero (i.e., flat) while those for PC$_\textrm{min}$ are significantly sloped. This is similar to the nature of PC relations for long period classical Cepheids ($P>10$d) in Galaxy, LMC, and SMC \citep{bhardwaj2014}. The similarity of the PC relations between W~Vir and classical Cepheids may be explained by their occupation of the same regions on the HR diagram and thus, equivalent $L/M$ ratios and effective temperatures. This observational evidence is thus in support of the HIF-stellar photosphere interaction theory. BL~Her stars, on the other hand, have a positive PC slope in the Bulge and the LMC and a flat (but negative) slope in the SMC at both minimum and maximum light. The nature of the sloped PC relations at both minimum and maximum light for BL~Her stars is similar to those for the short period classical Cepheids.

\section{Results from the theoretical aspect}

\begin{figure*}
\centering
\includegraphics[scale = 0.65]{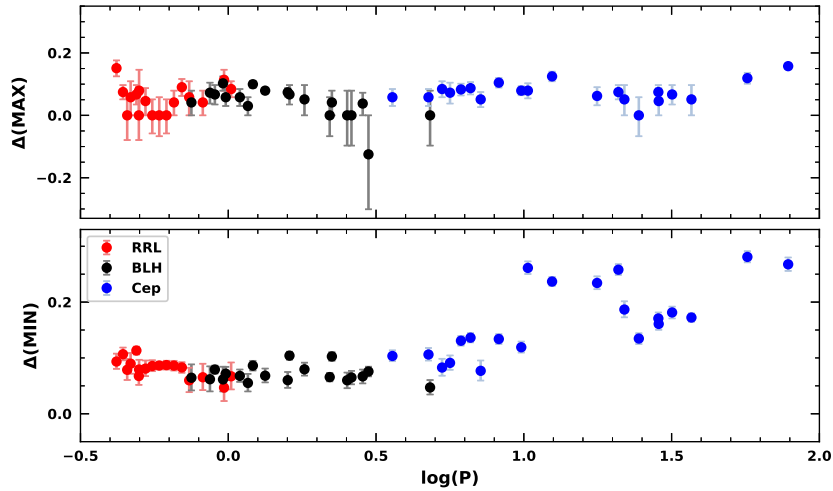}
\caption{The plots of distance ($\Delta$) between HIF and stellar photosphere as function of $\log(P)$ at maximum and minimum light.} 
\label{fig:HIF_Ph_distance} 
\end{figure*}

To study the HIF-photosphere interaction from a theoretical perspective, we compute RRL, BL~Her, and classical Cepheid models using the radial stellar pulsation code in MESA \citep{paxton2019}. We choose to use the convection parameter set A from Table~4 of \citet{paxton2019} which corresponds to the simplest convection model. For our analysis, we proceed with the models that have full-amplitude stable pulsations in the fundamental mode, an example of which is shown in Fig.~\ref{fig:lc}. The stellar photosphere is defined as the zone with optical depth of 2/3 while the HIF is the zone with the steepest gradient in temperature. The HIF-photosphere distance, $\Delta$ is then defined in terms of $Q=\log(1-M_r/M)$ \citep{kanbur2004b}. Fig.~\ref{fig:HIF_Ph_distance} displays the HIF-photosphere distance as a function of $\log(P)$ for the computed models at minimum and maximum light. For a particular model, we find that the HIF-photosphere distance is always smaller at maximum light. The top panel of Fig.~\ref{fig:HIF_Ph_distance} shows the HIF-photosphere distance of the different variable star types to be almost constant at maximum light. However, from the lower panel, we may clearly observe the smaller HIF-photosphere distance for the RRL models at minimum light (a situation of ``engaged'' HIF-photosphere) as opposed to the much larger HIF-photosphere distance for classical Cepheid models. BL~Her models show an intermediate behaviour to those from RRL and classical Cepheid models.

\section{Discussion and Conclusions}

We have analysed the PC relations of the different subclasses of T2Cs in the Galactic bulge and the Magellanic Clouds using OGLE-IV data. We find the W~Vir stars to have a flat PC slope at maximum light and a significantly sloped PC$_\textrm{min}$; a situation similar to long-period ($P>10$d) classical Cepheids \citep{bhardwaj2014}. On the other hand, BL~Her stars have sloped PC relations at both minimum and maximum light, similar to the short-period classical Cepheids. These results may be explained by the HIF-stellar photosphere interaction theory \citep{kanbur1996}. A lower $L/M$ ratio and/or a hotter effective temperature indicates smaller HIF-photosphere distance. This is true for W~Vir stars and classical Cepheids at maximum light when the temperature of the photosphere is the same as the temperature of the HIF, thereby resulting in a flat PC$_\textrm{max}$. RRLs always have an ``engaged'' HIF-photosphere. However, at minimum light, this occurs at a regime where Saha ionisation equilibrium is somewhat independent of the temperature, thereby resulting in a flat PC relation at minimum light \citep{das2018}. BL~Her stars occupy a region intermediate to the RRLs and classical Cepheids on the HR diagram, and this may explain the intermediate nature of their PC relations.

We have also computed RRL, BL~Her, and classical Cepheid models using MESA \citep{paxton2019} to theoretically test out the HIF-photosphere interaction theory by looking at their relative locations. We find that HIF-photosphere distance to be always smaller at maximum light, for a particular model. At minimum light, we find the HIF-photosphere distances for the classical Cepheid models to be much larger than that for the RRL models. The MESA computed models therefore support the HIF-photosphere interaction theory to a reasonable extent.

{\footnotesize \acknowledgements SD acknowledges the INSPIRE SRF provided by DST, Govt. of India vide Sanction Order No. DST/INSPIRE Fellowship/2016/IF160068 and the travel support provided by SERB, Govt. of India vide file number ITS/2019/004781 to attend the RR Lyrae/Cepheid 2019 Conference ``Frontiers of Classical Pulsators: Theory and Observations''. HPS and SMK thank the IUSSTF for funding the Indo-US virtual joint networked centre on ``Theoretical analyses of variable star light curves in the era of large surveys''. Funding for the Stellar Astrophysics Centre is provided by The Danish National Research Foundation (\#DNRF106). AB acknowledges research grant \#11850410434 awarded by the National Natural Science Foundation of China through the Research Fund for International Young Scientists, China Postdoctoral General Grant (2018M640018), and Peking University Strategic Partnership Fund awarded to Peking-Tokyo Joint-collaboration on Research in Astronomy and Astrophysics. The authors acknowledge the use of High Performance Computing facility Pegasus at IUCAA, Pune and the software MESA~r$11701$.}



\end{document}